\RequirePackage{fix-cm}
\documentclass[smallextended]{svjour3}       
\smartqed  
\usepackage{graphicx}
\usepackage{geometry}
\usepackage{mathrsfs}
\usepackage{amssymb}
\usepackage{amsmath}
\begin{document}

\title{Approach to Solving Quasiclassical Equations with Gauge Invariance}

\author{Priya Sharma}

\institute{Department of Physics,
              Indian Institute of Science,
              Bangalore 560012,
              India\\
              \email{shpriya@iisc.ac.in}}

\date{Received: date / Accepted: date}
\maketitle

\begin{abstract}
Quasiclassical equations with manifest gauge invariance are discussed in the context of unconventional singlet superconducting states in the static limit. Deviations of the quasiclassical propagator from its equilibrium solutions in the presence of magnetic fields and Hall terms are analysed in terms of a "{\it{small}}" parameter and a formulation developed to first order in "{\it{small}}". A modified quasiclassical propagator is defined to this order that is a solution of a new gauge-invariant Eilenberger-like equation with a normalisation condition. A Ricatti parametrisation with manifest gauge invariance is proposed. This theory is directly applicable to  homogenous $d$-wave order parameters in the presence of magnetic fields, such as in high-temperature superconductors.

\end{abstract}

\section{Introduction}
\label{intro}
The quasiclassical theory\cite{RainerSerene} is established as a powerful method to describe superconducting systems in the presence of external perturbations that vary slowly on the scale of the Fermi energy and Fermi wavelength. The central object of the theory is the quasiclassical Green's function for quasiparticles travelling with Fermi velocity, $v_F$ along ballistic trajectories defined by their direction $\hat{v}_F$. This propagator, represented as a Nambu matrix, is given by solutions of the Eilenberger equation\cite{Eilenberger}, along with a normalization condition that picks the physical solution. In the presence of magnetic fields, additional forces on the quasiparticles such as the Lorentz force appear as driving terms in the quasiclassical equation and have been included in a gauge-invariant formulation by Kita\cite{Kita}. The augmented Eilenberger equations have been applied to study vortex core charging, Hall currents and vortex lattices in superconductors \cite{Kita,Ueki,Kato}. 

The quasiclassical equation (Eilenberger as well as augmented versions) is a differential equation of the first order and usually, the solutions must be found numerically. In addition, the pair potential or the superconducting order parameter and self-energies (which are functions of the propagator) need to be solved for self-consistently. This is a nonlinear problem that is much simplified by a parametrization of the quasiclassical propagator in terms of coherence amplitudes that are solutions of a differential equation of Ricatti form \cite{Schopol}. This Ricatti parametrization transforms the Eilenberger equation for the matrix propagator to a scalar differential equation for the Ricatti amplitudes.  The solution then reduces to  that of an initial value problem to a scalar differential equation. It is known that the Riccati parametrization of the quasiclassical propagator leads to a stable and fast numerical method to solve the Eilenberger equations. 

In the presence of external magnetic fields, the quasiclassical propagator is found by solving the augmented Eilenberger equations. However, there is no simple normalization condition for the augmented quasiclassical propagator and physical solutions have to picked by imposing conservation laws and their physical behaviour at zero fields or large energies. In this paper, I derive a normalization condition for a modified propagator and a Ricatti parametrization for the augmented Eilenberger equations to leading order in a  {\it{"small"}} parameter in the static limit.
   
\section{Augmented Quasiclassical Theory in the Static Limit}

The Keldysh formalism is used to describe dynamical phenomena in quasiclassical theory.  The information about the spectrum of quasiparticles is carried by the Retarded and Advanced parts of the quasiclassical Green's function, $\hat{g}^{R,A}$. The quasiparticle distribution functions are described by the Keldysh part, $\hat{g}^K$. In the following section, I focus on the Retarded part and the derivation may be extended to Advanced and Keldysh parts later.
\vspace*{3mm}

\noindent
Kita\cite{Kita} derives a quasiclassical equation that retains a manifest gauge-invariance with respect to the space-time arguments of the quasiclassical Green's function, $\hat{g}$. In the static limit,  dropping the superscript $R$ for the Retarded part, this equation for $\hat{g}^R$ is :
\begin{equation}
\label{Kita}
[\varepsilon\hat{\tau}_3 - \hat{\sigma},\hat{g}]_{\circ} + i\,\vec{v}_F\cdot\eth_{\vec{R}}\hat{g} + (\frac{i}{2}\,\frac{e}{c}\,(\vec{v}_F\times\vec{H})\cdot\frac{\partial}{\partial\vec{p}} + \frac{i}{2}\,e\,\vec{v}_F\cdot\vec{E}\,\frac{\partial}{\partial\varepsilon})\,\{\hat{\tau}_3,\hat{g}\}_{\circ} = 0\,\,\,,
\end{equation}
where the gauge-transformed quasiclassical propagator for quasiparticles with momentum $\vec{p} = p_F\hat{p}$ and energy $\varepsilon$  at position $\vec{R}$ is $\hat{g}(\vec{p},\vec{R},\varepsilon)$. $\hat{\sigma}$ is the gauge-transformed quasiclassical self-energy. Quasiparticles moving with a velocity $\vec{v} = v_F\hat{p}$ are in a magnetic field $\vec{H} = \nabla\times\vec{\mathcal{A}}$ and electric field $\vec{E}$. $e,c$ are the constants electron charge and speed of light respectively. $\hat{\tau}_{1,2,3}$ are the Pauli matrices in Nambu space. The $\circ$-product is the folding product
\begin{equation}
\hat{a}\circ\hat{b}(\hat{p},\vec{R},\varepsilon,\varepsilon') = \int\nolimits\frac{d\varepsilon_1}{2\pi}\hat{a}(\hat{p},\vec{R},\varepsilon,\varepsilon_1)\hat{b}(\hat{p},\vec{R},\varepsilon_1,\varepsilon')\,\,\,.
\end{equation}
$\eth$ is defined \cite{Kita} as 
\begin{equation}
\eth \equiv \Big(\begin{array}{c c}
\frac{\partial}{\partial\vec{R}} & \frac{\partial}{\partial\vec{R}} - 2i\frac{e}{c}\mathcal{A}(\vec{R})\\
\frac{\partial}{\partial\vec{R}} + 2i\frac{e}{c}\mathcal{A}(\vec{R}) & \frac{\partial}{\partial\vec{R}}
\end{array}\Big)\,\,\,.
\end{equation}

\noindent
In the static limit, $\hat{g} = \hat{g}(\hat{p},\vec{R};\varepsilon)$ and the $\circ$ product reduces to matrix multiplication in Nambu indices. I will drop the $\circ$-product notation assuming matrix multiplication for the rest of this paper.
\vspace*{3mm}

\noindent
Equation(\ref{Kita}) is invariant under the gauge-transformation $\vec{\mathcal{A}} \rightarrow \vec{\mathcal{A}} + \nabla \chi$ ; $\Phi \rightarrow \Phi + \partial\chi/\partial t$ of the magnetic vector potential and electric potential respectively, with $\hat{g}\rightarrow e^{i\frac{e}{c}\chi\hat{\tau}_3}\hat{g}e^{-i\frac{e}{c}\chi\hat{\tau}_3}$ for $\chi = \chi(\vec{R})$. 
\noindent
Consider a system with a two-dimensional Fermi surface in the plane perpendicular to $\vec{H}$.  $\vec{v}_F = {v_F}_{\perp}$ and $(\vec{v}_F\times\vec{H})\cdot\partial/\partial p_{\perp} = 0$. Therefore,
\begin{equation}
\label{FS}
(\vec{v}_F\times\vec{H})\cdot\frac{\partial}{\partial \vec{p}} = v_F\,\frac{H}{p_F}\,\frac{\partial}{\partial \phi}\,\,\,,
\end{equation}
where $\phi$ is the angle in the plane of the Fermi surface i.e., $\vec{p} = p_F (\hat{p},\phi)$ and
\begin{equation}
\label{Larmor}
\frac{i}{2}\,\frac{e}{c}\,(\vec{v}_F\times\vec{H})\cdot\frac{\partial}{\partial \phi}\{\hat{\tau}_3,\hat{g}\} 
= i\omega_c\,\frac{\partial}{\partial \phi}\,\{\hat{\tau}_3,\hat{g}\}\,\,\,.
\end{equation} 
Here, $\omega_c$ is the Larmor frequency. Define an operator
\begin{equation}
\label{Omega}
\hat{\Omega} \equiv i\omega_c\frac{\partial}{\partial\phi} + \frac{i}{2}(\vec{v}_F\cdot\vec{E})\frac{\partial}{\partial\varepsilon}\,\,\,.
\end{equation}
Note that $\{\hat{\tau}_3,\hat{g}\}$ 
is diagonal in Nambu space.
Using equation(\ref{Larmor}), equation(\ref{Kita}) becomes
\begin{equation}
\label{Diag1}
[\varepsilon\hat{\tau}_3 - \hat{\sigma},\hat{g}]_{ll} + i\vec{v}_F\cdot\nabla g_{ll} + \hat{\Omega}\{\hat{\tau}_3,\hat{g}\}_{ll} = 0 \,\,\,\,\,\,\,;\,\,\,\,\,\,\, l = 1,2
\end{equation}
for the Nambu diagonal part (subscripts denoting Nambu indices) and 
\begin{equation}
\label{OffDiag1}
[\varepsilon\hat{\tau}_3 - \hat{\sigma},\hat{g}]_{jk} + i\vec{v}_F\cdot\nabla g_{jk} \pm 2\frac{e}{c}\,\vec{v}_F\cdot\vec{\mathcal{A}}\,g_{jk} = 0 \,\,\,\,\,\,\,;\,\,\,\,\,\,\, j\neq k \,; \, jk = 12, 21
\end{equation} 
for the Nambu off-diagonal part. 

\vspace*{3mm}
\noindent
The characterisic equation for the Eilenberger equation has {\it{characteristic curves}} that are straight line trajectories parallel to $\hat{v}_F$. For the Kita equation(\ref{Kita}), the characteristic curve 
is a circular orbit of radius $R_c = \pm\frac{v_F}{2\omega_c} $, the cyclotron radius for a given magnetic field $\vec{H}$. 
%
%
For electrons in cuprates, 
$R_c \sim 10^{-6} m \ll k_F^{-1}$ for fields of $H\sim 1T$.
\noindent
$R_c\propto H^{-1}$ and the orbit radius get larger as  the magnetic field gets smaller. As $H \rightarrow 0$, $ R_c \rightarrow \infty$ and the trajectories are straight lines as in the Eilenberger case. The augmented quasiclassical propagator is the solution to the Kita equation along circular trajectories of radius set by the Larmor frequency.

\section{Expansion in a Small Parameter}

In the absence of magnetic fields,
\begin{equation}
[\varepsilon\hat{\tau}_3 - \hat{\sigma}_0,\hat{g}_0] + i \vec{v}_F\cdot\nabla{\hat{g}}_0 = 0
\end{equation}
recovering the Eilenberger equation. For a  cuprate-like order parameter and  singlet state, the spin structure of $\hat{g},\hat{\sigma},\hat{\Delta}$ is trivial and spin scalar. For the  equilibrium case,
\begin{equation}
\label{eqmnorm}
\hat{\sigma}_0 = \Big(\begin{array}{c c} 0 & \Delta\\ \underline{\Delta} & 0\end{array}\Big) \equiv\hat{\Delta} \,\,\,\& \,\,\, \hat{g}_0\circ\hat{g}_0 = -\pi^2\hat{1},
\end{equation}
the normalization condition. $\hat{1}$ is a unit vector in Nambu space. $\Delta$ and $\underline{\Delta}$ are related by particle-hole conjugation $\Delta(\vec{p},\vec{R},\varepsilon) = \underline{\Delta}(-\vec{p},\vec{R},-\varepsilon)$. The equilibrium solution is 
\begin{equation}
\label{eqm}
\hat{g}_0 = -\pi\frac{\varepsilon\hat{\tau}_3 - \hat{\Delta}}{\sqrt{|\hat{\Delta}|^2 - \varepsilon^2}} \equiv \zeta(\varepsilon\hat{\tau}_3 - \hat{\Delta})\,\,\,.
\end{equation} 
The self-energies vary on the scale of the gap(magnitude of the order parameter) $\sim\Delta$. 
For cuprates, $\Delta\sim 10 meV$. Now, if we assume that the vector potential varies slowly on the scale of the Fermi wavelength, viz.. the quasiclassical length scale,
then for the cuprates $2\frac{e}{c}\frac{(\vec{v}_F\cdot\vec{\mathcal{A}})}{\Delta} \ll 1$ and $\frac{\hbar\omega_c}{\Delta} \ll 1$. It follows that for the cuprates with 2D Fermi surface as assumed at the outset, the driving term in the Kita equation ($\propto \hat{\Omega}$) and the Kita derivative term ($\propto 2\frac{e}{c}\mathcal{A}$) can be treated as {\it{"small"}} and $\hat{g}$ expanded in this {\it{"small"}} parameter for fields $\leq 1T$. {\it{"small"}} $\sim \frac{\hbar\omega_c}{\Delta}\,\,,\,\,2\frac{e}{c}\frac{\vec{v}_F\cdot\vec{\mathcal{A}}}{\Delta}$. Let
\begin{equation}
\label{glinear}
\hat{g} = \hat{g}_0({\it{small}}^0) + \hat{g}_1({\it{small}}^1)\,\,\,\,\,.
\end{equation}
Linearizing equation(\ref{Diag1} and \ref{OffDiag1}),  the diagonal and off-diagonal parts to $\mathcal{O}({\it{small}}^0)$ give the Eilenberger equation in the absence of fields. To $\mathcal{O}({\it{small}}^1)$, 
\begin{equation}
[\varepsilon\hat{\tau}_3 - \hat{\Delta},\hat{g}_1]_{ii} - [\hat{\sigma}_1,\hat{g}_0]_{ii} + i\vec{v}_F\cdot\nabla(\hat{g}_1)_{ii} + \hat{\Omega}\{\hat{\tau}_3,\hat{g}_0\}_{ii} = 0\,\,\,\,.
\end{equation}
Here $\hat{\sigma} = \hat{\sigma}_0 ({\it{small}}^0) + \hat{\sigma}_1({\it{small}}^1) = \hat{\Delta} + \hat{\sigma}_1$. With the definition in equation(\ref{eqm}), $\hat{\Omega}(\{\hat{\tau}_3,\hat{g}_0\}) = 2\hat{\Omega}(\zeta\varepsilon)\hat{1}$.
Therefore, to $\mathcal{O}({\it{small}}^1)$, the diagonal part of the equation for $\hat{g}$ is 
\begin{equation}
\label{OsmallDiag}
[\varepsilon\hat{\tau}_3 - \hat{\Delta},\hat{g}_1]_{ii} - [\hat{\sigma}_1,\hat{g}_0]_{ii} + (i\vec{v}_F\cdot\nabla\hat{g}_1)_{ii} + 2\hat{\Omega}(\zeta\varepsilon) = 0
\end{equation}
and the off-diagonal part (to $\mathcal{O}({\it{small}}^1)$) is 
\begin{equation}
\label{OsmallOffDiag}
[\varepsilon\hat{\tau}_3 - \hat{\Delta},\hat{g}_1]_{jk} - [\hat{\sigma}_1,\hat{g}_0]_{jk} + (i\vec{v}_F\cdot\nabla\hat{g}_1)_{jk} \pm \frac {2e}{c}(\vec{v}_F\cdot\vec{\mathcal{A}})\hat{g}_0\mid_{jk} = 0\,\,\,.
\end{equation}

\noindent
The natural question arises as to the gauge-invariance of these equations to $\mathcal{O}({\it{small}}^1)$.
The gauge-invariant equation(\ref{Kita}) derived by Kita[1] is an equation that remains unchanged under the transformation $\vec{\mathcal{A}} \rightarrow \vec{\mathcal{A}} + \nabla\chi$; $\Phi \rightarrow \Phi + \frac{\partial\chi}{\partial t}$ ; $\hat{g} \rightarrow e^{i\frac{e}{c}\chi\hat{\tau}_3}\hat{g}e^{-i\frac{e}{c}\chi\hat{\tau}_3}$ with $\chi = \chi(\vec{R})$. In other words, the solution to the gauge-invariant equation does not depend upon $\chi(\vec{R})$. For any $\chi(\vec{R})$, this solution transforms as above. Now, consider  an equation to $\mathcal{O}({\it{small}}^1)$. If the equation remains unchanged under a known transformation for arbitrary $\chi(\vec{R})$ to $\mathcal{O}({\it{small}}^1)$, i.e., as long as only terms to $\mathcal{O}({\it{small}}^1)$ are considered, then it can be said to be gauge-invariant to $\mathcal{O}({\it{small}}^1)$. Equations(\ref{OsmallDiag} and \ref{OsmallOffDiag}) are gauge-invariant precisely in this sense. I now derive an equation for a new $\hat{\mathscr{G}}$, that is gauge-invariant to $\mathcal{O}({\it{small}}^1)$ in the sense just described.

\section*{Equation for a Modified $\hat{\mathscr{G}}$}
$\hat{g}$  is given by a gauge-invariant equation(\ref{Kita}) with a normalisation $\hat{\nu} \equiv \hat{g}\circ\hat{g}$ given by [1]
\begin{equation}
\label{nuKita}
[\varepsilon\hat{\tau}_3 - \hat{\sigma},\hat{\nu}] + i\vec{v}_F\cdot\eth_{\vec{R}}\hat{\nu} = \{\hat{g}, -\hat{\Omega}(\{\hat{\tau}_3,\hat{g}\})\}\,\,\,.
\end{equation}
Now, to $\mathcal{O}({\it{small}}^1)$, the driving term is given by $-4\hat{\Omega}(\zeta\varepsilon)\hat{g}_0$.
Linearizing $\hat{\nu}$, let
\begin{eqnarray}
\label{nulinear}
\hat{\nu} &=& \hat{\nu}_0({\it{small}}^0) + \hat{\nu}_1({\it{small}}^1) \\
\nonumber
\hat{\nu}_0 &=& \hat{g}_0\circ\hat{g}_0 = -\pi^2\hat{1} \,\,\,;\,\,\, \hat{\nu}_1 = \hat{g}_0\circ\hat{g}_1 + \hat{g}_1\circ\hat{g}_0
\end{eqnarray}
using the equilibrium condition(\ref{eqmnorm}) and keeping terms to $\mathcal{O}({\it{small}}^1)$ using equation(\ref{glinear}). Equation(\ref{nuKita}) is trivially satisfied to $\mathcal{O}({\it{small}}^0)$. To $\mathcal{O}({\it{small}}^1)$,
\begin{equation}
\label{nu1}
[\varepsilon\hat{\tau}_3 - \hat{\Delta},\hat{\nu}_1] - [\hat{\sigma}_1,\hat{\nu}_0] + i\vec{v}_F\cdot\eth(\hat{\nu}_0 + \hat{\nu}_1) = -4\hat{\Omega}(\zeta\varepsilon)\hat{g}_0\,\,\,.
\end{equation}
Using equation(\ref{eqm}), to $\mathcal{O}({\it{small}}^1)$,
\begin{equation}
i\vec{v}_F\cdot\eth\hat{\nu} \rightarrow i\vec{v}_F\cdot\nabla\hat{\nu}_1 \pm 2\frac{e}{c}(\vec{v}_F\cdot\vec{\mathcal{A}})\hat{\nu}_0\mid_{12,21} = i\vec{v}_F\cdot\nabla\hat{\nu}_1\,\,\,.
\end{equation}
Equation(\ref{nu1}) simplifies to
\begin{equation}
\zeta^{-1}[\hat{g}_0,\hat{\nu}_1] + i\vec{v}_F\cdot\nabla\hat{\nu}_1 = -4\hat{\Omega}(\zeta\varepsilon)\hat{g}_0\,\,\,.
\end{equation}
Multiplying from the left and right by $\hat{g}_0$ and adding the two resulting equations, I get
\begin{eqnarray}
\label{defineN}
&\,&i\vec{v}_F\cdot\nabla\{\hat{g}_0,\hat{\nu}_1\} = 8\pi^2\hat{\Omega}(\zeta\varepsilon)\\
\nonumber
&\Rightarrow& \{\hat{g}_0,\hat{\nu}_1\} = \frac{1}{iv_F}\oint ds\,8\pi^2\hat{\Omega}(\zeta\varepsilon) = -i\frac{8\pi^3}{\omega_c}\hat{\Omega}(\zeta\varepsilon) \equiv \mathcal{N}(\varepsilon,\phi,s)\,\,\,,
\end{eqnarray}
where I use the equilibrium normalization(\ref{eqmnorm}), $\hat{s}\parallel\hat{v}_F$ is a coordinate along a circular trajectory along which the integral denoted by $\oint$ is performed. Here, I assume the equilibrium solution to be the homogenous solution. While this assumption is valid only in the limit of the homogenous state, viz., far away from vortices for applied fields $H \ll H_{c2}$, it leads to an elegant result that effects the Ricatti parametrization, as I show shortly. Now, define
\begin{equation}
\label{defineg0}
\hat{g}_1 = \mathcal{G}_0 \hat{1} + \delta\hat{g}\,\,\,,
\end{equation}
where $\delta\hat{g}\propto\hat{\tau}_1,\hat{\tau}_2,\hat{\tau}_3$ only. Then $\hat{\nu}_1$ as given by equation(\ref{nulinear}) is expressed as
\begin{equation}
\hat{\nu}_1 = 2\mathcal{G}_0\hat{g}_0 + \{\hat{g}_0,\delta\hat{g}\}\,\,\,,
\end{equation}
and
\begin{equation}
\{\hat{g}_0,\hat{\nu}_1\} = -2\pi^2\hat{g}_1 + 2\hat{g}_0\circ\hat{g}_1\circ\hat{g}_0\,\,\,.
\end{equation}
Using the known $\hat{g}_0$ (\ref{eqm}), 
\begin{eqnarray}
\label{G0}
\mathcal{N}(\varepsilon,\phi,s) &=& -2\pi^2\mathcal{G}_0 - 2\pi^2\mathcal{G}_0 = -4\pi^2\mathcal{G}_0
\\
\nonumber
\Rightarrow \mathcal{G}_0 &=& \frac{-1}{4\pi^2}\mathcal{N}(\varepsilon,\phi,s)\,\,\,!
\end{eqnarray}
Using the definition(\ref{defineg0}), 
the normalization condition for $\hat{g}_0$, and the result above, some further algebra gives
\begin{equation}
\{\hat{g}_0,\delta\hat{g}\} = 0 \,\,\,.
\end{equation}
I now define a new $\hat{\mathscr{G}}$  thus 
\begin{equation}
\hat{g} = \hat{g}_0 + \mathcal{G}_0\hat{1} + \delta\hat{g} = \mathcal{G}_0\hat{1} + \hat{\mathscr{G}}
\end{equation}
and evaluate $\hat{\mathscr{G}}\circ\hat{\mathscr{G}}$ to find
\begin{equation}
\hat{\mathscr{G}}\circ\hat{\mathscr{G}} \rightarrow -\pi^2 + \{\hat{g}_0,\delta\hat{g}\} = -\pi^2\hat{1}\,\,\,!
\end{equation}
to $\mathcal{O}({\it{small}}^1)$.
\vspace*{3mm}

\noindent
Summarising, I separate the part of $\hat{g}$ that is $\propto\hat{1}$, viz., $\hat{g} = \mathcal{G}_0\hat{1} + \hat{\mathscr{G}}$ and define a modified  $\hat{\mathscr{G}}$ that is traceless and satisfies the usual normalization condition.
\vspace*{3mm}

\noindent
$\hat{\mathscr{G}}$ is given by an equation :
\begin{equation}
\label{newGequation}
[\varepsilon\hat{\tau}_3 - \hat{\sigma},\hat{\mathscr{G}}] + i\vec{v}_F\cdot\eth\mathcal{G}_0\hat{1} + i\vec{v}_F\cdot\eth\hat{\mathscr{G}} + \hat{\Omega}(\{\hat{\tau}_3,\mathcal{G}_0\hat{1} + \hat{\mathscr{G}}\}) = 0
\end{equation}
with normalization condition
\begin{equation}
\label{newnorm}
\hat{\mathscr{G}}\circ\hat{\mathscr{G}} = -\pi^2\hat{1}\,\,\,.
\end{equation}

\noindent
Each term in equation(\ref{newGequation}) can be rigorously shown to be invariant under the transformation $\vec{\mathcal{A}} \rightarrow \vec{\mathcal{A}} + \nabla\chi(\vec{R})$ ; $\hat{g} \rightarrow e^{i\frac{e}{c}\chi(\vec{R})\hat{\tau}_3}\,\hat{\mathscr{G}}\,e^{-i\frac{e}{c}\chi(\vec{R})\hat{\tau}_3}$, i.e., equation(\ref{newGequation}) is a gauge-invariant equation for $\hat{\mathscr{G}}$.

\section*{Parametrization of $\hat{\mathscr{G}}$}

\noindent
With the normalization condition (\ref{newnorm}), $\hat{\mathscr{G}}$ can be parametrized thus:
\begin{equation}
\label{parameter}
\hat{\mathscr{G}} = \frac{-i\pi}{1 - \gamma\tilde{\gamma}}\Big(\begin{array}{c c}
1 + \gamma\tilde{\gamma} & 2\gamma\\ 
-2\tilde{\gamma} & -1-\gamma\tilde{\gamma}\end{array}\Big)\,\,\,.
\end{equation}
To $\mathcal{O}({\it{small}}^0)$, $\hat{\mathscr{G}}\rightarrow\hat{g}_0$ and $\gamma$,$\tilde{\gamma}$ are the usual Ricatti amplitudes. To $\mathcal{O}({\it{small}}^1)$, denote $\varepsilon\hat{\tau}_3 - \hat{\sigma} = \tilde{\varepsilon}\hat{\tau}_3 - \hat{\Delta}'$. 
Using the parametrization(\ref{parameter}) in equation(\ref{newGequation}), 
I  derive a set of equation for the parameters $\gamma,\tilde{\gamma}$ that are Ricatti equations :
\begin{eqnarray}
\label{Ricatti}
2(\tilde{\varepsilon} + \frac{e}{c}(\vec{v}_F\cdot\vec{\mathcal{A}}))\gamma + \Delta' - \tilde{\Delta}'\gamma^2 + i\vec{v}_F\cdot\nabla\gamma &=& 0
\nonumber
\\
2(\tilde{\varepsilon} - \frac{e}{c}(\vec{v}_F\cdot\vec{\mathcal{A}}))\tilde{\gamma} - \tilde{\Delta}' + \Delta'\tilde{\gamma}^2 - i\vec{v}_F\cdot\nabla\tilde{\gamma} &=& 0
\end{eqnarray}
These Ricatti equations are invariant under the transformation $\gamma\rightarrow\gamma \,e^{2i\frac{e}{c}\chi(\vec{R})}$ ; $\tilde{\gamma} \rightarrow\tilde{\gamma}e^{-2i\frac{e}{c}\chi(\vec{R})}$ ; $\Delta' \rightarrow\Delta'e^{2i\frac{e}{c}\chi(\vec{R})}$; $\tilde{\Delta}' \rightarrow \tilde{\Delta}'e^{-2i\frac{e}{c}\chi(\vec{R})}$ ; $\vec{\mathcal{A}}\rightarrow\vec{\mathcal{A}} + \nabla\chi(\vec{R})$. $\Delta'$, $\tilde{\Delta}'$ transform as the off-diagonal parts of $\hat{\mathscr{G}}$ transform. Equations(\ref{Ricatti}) can be solved for $\gamma$, $\tilde{\gamma}$. $\hat{\mathscr{G}}$ can be constructed from $\gamma$, $\tilde{\gamma}$. The full gauge-transformed $\hat{g}$ to $\mathcal{O}({\it{small}}^1)$ is given by $\hat{g} = \mathcal{G}_0\hat{1} + \hat{\mathscr{G}}$. 

\section{Discussion}

$\mathcal{G}_0$ contributes to the Retarded and Advanced parts of $\hat{g}$ and drops off from the Keldysh part. In physical terms, $\mathcal{G}_0$ contributes to the spectral density of states. The distribution functions are not affected by $\mathcal{G}_0$. More generally, the latter breaks the particle-hole symmetry of the quasiparticle propagator $\hat{g}$ in the presence of magnetic fields. To leading order in a {\it{small}} parameter, separation of this part renders $\mathscr{G}$  traceless and this can be Ricatti parametrised for homogenous ground states. While the Ricatti equations in the form equation(\ref{Ricatti}) are the usual ones applied widely to superconductors in magnetic fields, I have derived them as a natural outcome of a manifest gauge transformation that generates the driving terms of the quasiclassical equation in the presence of magnetic fields. These  gauge-transformations are usually applied as an ansatz from physical arguments. This provides a rigorous derivation of these transformations as well as the use of the Ricatti equation in the presence of magnetic field terms  within the quasiclassical method.  

\begin{acknowledgements}
I acknowledge financial support from the Department of Science and Technology, Government of India : Scheme No. SR/WOS-A/PM-4/2016. 
\end{acknowledgements}



\end{document}